\documentclass[peerreview,11pt,draftclsnofoot,onecolumn,a4paper]{IEEEtran}
\usepackage{epsfig}
\usepackage[cmex10]{amsmath}
\usepackage{amsfonts,amssymb}

\usepackage[latin1]{inputenc}
\usepackage{algorithm}
\usepackage{algorithmic}
\begin{document}
\title{Distributed Arithmetic Coding for the Slepian-Wolf problem}

\author{Marco Grangetto,~\IEEEmembership{Member,~IEEE,}
Enrico~Magli,~\IEEEmembership{Senior Member,~IEEE,} Gabriella Olmo,
~\IEEEmembership{Senior Member,~IEEE}
\thanks{M. Grangetto is with Dip. di Informatica,
Universit\`a degli Studi di Torino, Corso Svizzera 185 - 10149
Torino - ITALY - Ph.: +39-011-6706711 - FAX: +39-011-751603 -
E-mail: {\tt marco.grangetto@di.unito.it} }
\thanks{E. Magli and G. Olmo are with  Dip. di Elettronica,
Politecnico di Torino, Corso Duca degli Abruzzi 24 - 10129 Torino -
Italy - Ph.: +39-011-5644195 - FAX: +39-011-5644099 - E-mail: {\tt
enrico.magli(gabriella.olmo)@polito.it}. Corresponding author:
Enrico Magli.}\\ $\;$ \\EDICS: SEN-DCSC, SPC-CODC}

\markboth{IEEE Transactions on Signal Processing (resubmitted
November 2008)}{Shell \MakeLowercase{\textit{et al.}}: Bare Demo of
IEEEtran.cls for Journals}

\maketitle

\begin{abstract}
Distributed source coding schemes are typically based on the use of
channels codes as source codes. In this paper we propose a new
paradigm, named ``distributed arithmetic coding", which extends
arithmetic codes to the distributed case employing sequential
decoding aided by the side information. In particular, we introduce
a distributed binary arithmetic coder for the Slepian-Wolf coding
problem, along with a joint decoder. The proposed scheme can be
applied to two sources in both the asymmetric mode, wherein one
source acts as side information, and the symmetric mode, wherein
both sources are coded with ambiguity, at any combination of
achievable rates. Distributed arithmetic coding provides several
advantages over existing Slepian-Wolf coders, especially good
performance at small block lengths, and the ability to incorporate
arbitrary source models in the encoding process, e.g., context-based
statistical models, in much the same way as a classical arithmetic
coder. We have compared the performance of distributed arithmetic
coding with turbo codes and low-density parity-check codes, and
found that the proposed approach is very competitive.
\end{abstract}

\begin{keywords}
Distributed source coding, arithmetic coding, Slepian-Wolf coding,
Wyner-Ziv coding, compression, turbo codes, LDPC codes.
\end{keywords}

\IEEEpeerreviewmaketitle

\section{Introduction and background}
\label{sec:intro}

In recent years, distributed source coding (DSC) has received an
increasing attention from the signal processing community. DSC
considers a situation in which two (or more) statistically dependent
sources $X$ and $Y$ must be encoded by separate encoders that are
not allowed to talk to each other. Performing separate lossless
compression may seem less efficient than joint encoding. However,
DSC theory proves that, under certain assumptions, separate encoding
is optimal, provided that the sources are decoded jointly
\cite{slepian}. For example, with two sources it is possible to
perform ``standard" encoding of the first source (called {\em side
information}) at a rate equal to its entropy, and ``conditional"
encoding of the second one at a rate lower than its entropy, with no
information about the first source available at the second encoder;
we refer to this as ``asymmetric" Slepian-Wolf (S-W) problem.
Alternatively, both sources can be encoded at a rate smaller than
their respective entropy, and decoded jointly, which we refer to as
``symmetric" S-W coding.

DSC theory also encompasses lossy compression \cite{wyner}; it has
been shown that, under certain conditions, there is no performance
loss in using DSC \cite{wyner,duality}, and that possible losses are
bounded below 0.5 bit per sample (bps) for quadratic distortion
metric \cite{zamir}. In practice, lossy DSC is typically implemented
using a quantizer followed by lossless DSC, while the decoder
consists of the joint decoder followed by a joint dequantizer.
Lossless and lossy DSC have several potential applications, e.g.,
coding for non co-located sources such as sensor networks,
distributed video coding \cite{prism,aaron,magli_mmsp05,discover},
layered video coding \cite{xiong_tip,ortega_dsc}, error resilient
video coding \cite{sehgal}, and satellite image coding
\cite{magli_dsc,ortega_sp}, just to mention a few. The interested
reader is referred to \cite{xiong_spm} for an excellent tutorial.

Traditional entropy coding of an information source can be performed
using one out of many available methods, the most popular being
arithmetic coding (AC) and Huffman coding. ``Conditional" (i.e.,
DSC) coders are typically implemented using channel codes, by
representing the source using the syndrome or the parity bits of a
suitable channel code of given rate. The syndrome identifies sets of
codewords (``cosets") with maximum distance properties, so that
decoding an ambiguous description of a source at a rate less than
its entropy (given the side information) incurs minimum error
probability. If the correlation between $X$ and $Y$ can be modeled
as a ``virtual" channel described as $X=Y+W$, with $W$ an additive
noise process, a good channel code for that transmission problem is
also expected to be a good S-W source code \cite{duality}.

Regarding asymmetric S-W coding, the first practical technique has
been described in \cite{discus}, and employs trellis codes.
Recently, more powerful channel codes such as turbo codes have been
proposed in \cite{aaron,frias,xiong_turbo}, and low-density
parity-check (LDPC) \cite{Gallager} codes have been used in
\cite{Xiong2,xiong_tcq,varodayan}. Turbo and LDPC codes can get
extremely close to channel capacity, although they require the block
size to be rather large. Note that the constituent codes of
turbo-codes are convolutional codes, hence the syndrome is difficult
to compute. In \cite{aaron} the cosets are formed by all messages
that produce the same parity bits, even though this approach is
somewhat suboptimal \cite{xiong_turbo}, since the geometrical
properties of these cosets are not as good as those of
syndrome-based coding. In \cite{blum2} a syndrome former is used to
deal with this problem. Multilevel codes have also be addressed; in
\cite{ramchandran_multi} trellis codes are extended to multilevel
sources, whereas in \cite{xiong_tsp} a similar approach is proposed
for LDPC codes.

Besides techniques based on channel coding, a few authors have also
investigated the use of source coders for DSC. This is motivated by the fact
that existing source coders obviously exhibit nice compression features that
should be retained in a DSC coder, such as the ability to employ flexible and
adaptive probability models, and low encoding complexity. In \cite{rose_dsc}
the problem of designing a variable-length DSC coder is addressed; it is shown
that the problem of designing a zero-error such coder is NP-hard. In
\cite{effros} a similar approach is followed; the authors consider the problem
of designing Huffman and arithmetic DSC coders for multilevel sources with zero
or almost-zero error probability. The idea is that, if the joint density of the
source and the side information satisfies certain conditions, the same codeword
(or the same interval for the AC process) can be associated to multiple
symbols. This approach leads to an encoder with a complex modeling stage
(NP-hard for the optimal code, though suboptimal polynomial-time algorithms are
provided in \cite{effros}), while the decoding process resembles a classical
arithmetic decoder.

As for symmetric S-W codes, a few techniques have been recently
proposed.
A symmetric code can be obtained from an asymmetric one through time
sharing, whereby the two sources alternatively take the role of the
source and the side information; however, current DSC coders cannot
easily accommodate this approach. Syndrome-based channel code
partitioning has been introduced in \cite{ramcha_partitioning}, and
extended in \cite{xiong_tit} to systematic codes. A similar
technique is described in \cite{dragotti_comlet}, encompassing
non-systematic codes. Syndrome formers have also been proposed for
symmetric S-W coding \cite{tiffany}. Moreover, techniques based on
the use of parity bits can also be employed, as they can typically
provide rate compatibility. A practical code has been proposed in
\cite{frias} using two turbo codes that are decoded jointly,
achieving the equal rate point; in \cite{frias_sp} an algorithm is
introduced that employs turbo codes to achieve arbitrary rate
splitting. Symmetric S-W codes based on LDPC codes have also been
developed \cite{fekri,guillemot_ldpc}.

Although several near-optimal DSC coders have been designed for
simple ideal sources (e.g., binary and Gaussian sources), the
applications of practical DSC schemes to realistic signals typically
incurs the following problems.

\begin{itemize}
\item Channel codes get very close to capacity only for very large data
blocks (typically in excess of $10^5$ symbols). In many
applications, however, the basic units to be encoded are of the
order of a few hundreds to a few thousands symbols. For such block
lengths, channel codes have good but not optimal performance.
\item The symbols contained in a block are expected to follow a
stationary statistical distribution. However, typical real-world
sources are not stationary. This calls for either the use of short
blocks, which weakens the performance of the S-W coder, or the
estimation of conditional probabilities over contexts, which cannot
be easily accommodated by existing S-W coders.
\item When the sources are strongly correlated (i.e., in the most favorable
case), very high-rate channel codes are needed (e.g.,
rate-$\frac{99}{100}$ codes). However, capacity-achieving channel
codes are often not very efficient at high rate.
\item In those applications where DSC is used to limit the encoder
complexity, it should be noted that the complexity of existing S-W
coders is not negligible, and often higher than that of existing
non-DSC coders. This seriously weakens the benefits of DSC.
\item Upgrading an existing compression algorithm like JPEG 2000 or
H.264/AVC to provide DSC functionalities requires at least to
redesign the entropy coding stage, adopting one of the existing DSC
schemes.
\end{itemize}

Among these issues, the block length is particularly important.
While it has been shown that, on ideal sources with very large block
length, the performance of some practical DSC coders can be as close
as 0.09 bits to the theoretical limit \cite{xiong_spm}, so far DSC
of real-world data has fallen short of its expectations, one reason
being the necessity to employ much smaller blocks. For example, the
PRISM video coder \cite{prism} encodes each macroblock
independently, with a block length of 256 samples. For the coder in
\cite{aaron}, the block length is equal to the number of 8x8 blocks
in one picture (1584 for the CIF format). The performance of both
coders is rather far from optimal, highlighting the need of DSC
coders for realistic block lengths.

A solution to this problem has been introduced in \cite{dac_comlet},
where an extension of AC, named distributed arithmetic coding (DAC),
has been proposed for asymmetric S-W coding. Moreover, in
\cite{tsdac} DAC has been extended to the case of symmetric S-W
coding of two sources at the same rate (i.e., the mid-point of the
S-W rate region). DAC and its decoding process do not currently have
a rigorous mathematical theory that proves they can asymptotically
achieve the S-W rate region; such theory is very difficult to
develop because of the non-linearity of AC. However, DAC is a
practical algorithm that was shown in \cite{dac_comlet} to
outperform other existing distributed coders. In this paper, we
build on the results presented in \cite{dac_comlet}, providing
several new contributions. For asymmetric coding, we focus on i.i.d.
sources as these are often found in many DSC applications; for
example, in transform-domain distributed video coding, DAC could be
applied to the bit-planes of transform coefficients, which can be
modeled as i.i.d. We optimize the DAC using an improved encoder
termination procedure, and we investigate the rate allocation
problem, i.e., how to optimally select the encoding parameters to
achieve a desired target rate. We evaluate the performance of this
new design comparing it with turbo and LDPC codes, including the
case of extremely correlated sources with highly skewed
probabilities. This is of interest in multimedia applications
because the most significant bit-planes of the transform
coefficients of an image or video sequence are almost always equal
to zero, and are strongly correlated with the side information. For
symmetric coding, we extend our previous work in \cite{tsdac} by
introducing DAC encoding and rate allocation procedures that allow
to encode an arbitrary number of sources with arbitrary combination
of rates. We develop and test the decoder for two sources.

Finally, it should be noted that an asymmetric DAC scheme has been
independently and concurrently developed in \cite{guillemot_dac}
using quasi-arithmetic codes. Quasi-arithmetic codes are a
low-complexity approximation to arithmetic codes, providing smaller
encoding and decoding complexity \cite{howard}. These codes allow
the interval endpoints to be only a finite set of points. While this
yields suboptimal compression performance, it makes the arithmetic
coder a finite state machine, simplifying the decoding process with
side information.

This paper is organized as follows. In Sect. \ref{sec:dac_decoder}
we describe the DAC encoding process for the asymmetric case, in
Sect. \ref{sec:dac_decoder} we describe the DAC decoder, and in
Sect. \ref{sec:rate_sel} we study the rate allocation and parameter
selection problem. In Sect. \ref{sec:symm} we describe the DAC
encoder, decoder and rate allocator for the symmetric case. In Sect.
\ref{sec:results_asymm} and \ref{sec:results_symm} we report the DAC
performance evaluation results in the asymmetric and symmetric case
respectively. Finally, in Sect. \ref{sec:concl} we draw some
conclusions.

\section{Distributed arithmetic coding: asymmetric encoder}
\label{sec:dac_encoder}

Before describing the DAC encoder, it should be noted that the AC
process typically consists of a modeling stage and a coding stage.
The modeling stage has the purpose of computing the parameters of a
suitable statistical model of the source, in terms of the
probability that a given bit takes on value 0 or 1. This model can
be arbitrarily sophisticated, e.g., by using contexts, adaptive
probability estimation, and so forth. The coding stage takes the
probabilities as input, and implements the actual AC procedure,
which outputs a binary codeword describing the input sequence.

\begin{figure}[tb]
\centering
\includegraphics[width=8cm,clip]{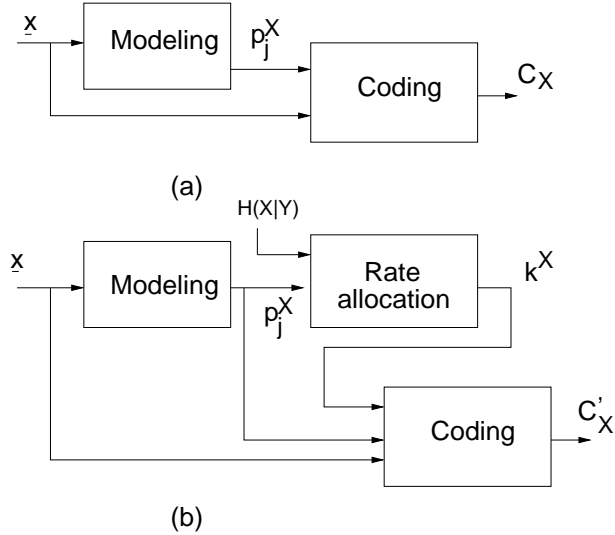}
\caption{Modeling, rate allocation and coding stage for (a)
classical AC, and (b) DAC.} \label{fig:modeling}
\end{figure}

Let $X$ be a binary memoryless source that emits a semi-infinite
sequence of random variables $X_i$, $i=0, 1, \ldots$, with
probabilities $p^X_{0}=P(X_i=0)$ and $p^X_{1}=P(X_i=1)$. We are
concerned with encoding the sequence $\underline{x}=[x_0, \ldots,
x_{N-1}]$ consisting in the first $N$ occurrences of this source.
The modeling and coding stages are shown in Fig.
\ref{fig:modeling}-a. The modeling stage takes as input the sequence
$\underline{x}$, and outputs an estimate of the probabilities
$p^X_{0}$ and $p^X_{1}$. The coding stage takes as input
$\underline{x}$, $p^X_{0}$ and $p^X_{1}$, and generates a codeword
$C_X$. The expected length of $C_X$ depends on $p^X_{0}$ and
$p^X_{1}$, and is determined once these probabilities are given.

In order to use the DAC, we consider two sources $X$ and $Y$, where
$Y$ is a binary memoryless source that emits random variables $Y_i$,
$i=0, 1, \ldots$, with probabilities $p^Y_{0}=P(Y_i=0)$ and
$p^Y_{1}=P(Y_i=1)$. The first $N$ occurrences of this source form
the side information $\underline{y}=[y_0, \ldots, y_{N-1}]$. We
assume that $X$ and $Y$ are i.i.d. sources, and that $X_i$ and $Y_i$
are statistically dependent for a given $i$. The entropy of $X$ is
defined as $H(X)=-\sum_{j=0}^1p^X_j \log_2 p^X_j$, and similarly for
$Y$. The conditional entropy of $X$ given $Y$ is defined as
$H(X|Y)=-\sum_{j=0}^1 \sum_{k=0}^1 P(X_i=j,Y_i=k) \log_2
P(X_i=j|Y_i=k)$.

For DAC, three blocks can be identified, as in Fig.
\ref{fig:modeling}-b, namely the modeling, rate allocation, and
coding stages. The modeling stage is exactly the same as in the
classical AC. The coding stage will be described in Sect.
\ref{sec:dac_enc}; it takes as inputs $\underline{x}$, the
probabilities $p^X_{0}$ and $p^X_{1}$, and the parameter $k^X$, and
outputs a codeword $C'_X$. Unlike a classical AC, where the expected
rate is function of the source probabilities, and hence cannot be
selected {\em a priori}, the DAC allows to select any desired rate
not larger than the expected rate of a classical AC. This is very
important, since in a DSC setting the rate for $\underline{x}$
should depend not only on how much ``compressible" the source is,
but also on how much correlated $X_i$ and $Y_i$ are. For this
reason, in DAC we also have a rate allocation stage that takes as
input the probabilities $p^X_{0}$ and $p^X_{1}$ and the conditional
entropy $H(X|Y)$, and outputs a parameter $k^X$ that drives the DAC
coding stage to achieve the desired target rate.

In this paper we deal with the coding and rate allocation stages,
and assume that the input probabilities $p^X_{0}$, $p^X_{1}$ and
conditional entropy $H(X|Y)$ are known {\em a priori}. This allows
us to focus on the distributed coding aspects of the proposed
scheme, and, at the same time, keeps the scheme independent of the
modeling stage.


\subsection{Arithmetic coding}
\label{sec:ac}

We first review the classical AC coding process, as this sets the
stage for the description of the DAC encoder; an overview can be
found in \cite{cacm}. The binary AC process for $\underline{x}$ is
based on the probabilities $p^X_{0}$ and $p^X_{1}$, which are used
to partition the $[0,1)$ interval into sub-intervals associated to
possible occurrences of the input symbols. At initialization the
``current" interval is set to $I_0=[0,1)$. For each input symbol
$x_i$, the current interval $I_{i}$ is partitioned into two adjacent
sub-intervals of lengths $p^X_{0}|I_{i}|$ and $p^X_{1}|I_{i}|$,
where $|I_i|$ is the length of $I_i$. The sub-interval corresponding
to the actual value of $x_i$ is selected as the next current
interval $I_{i+1}$, and this procedure is repeated for the next
symbol. After all $N$ symbols have been processed, the sequence is
represented by the final interval $I_N$. The codeword $C_X$ can
consist in the binary representation of any number inside $I_N$
(e.g., the number in $I_N$ with the shortest binary representation),
and requires approximately $-\log_2 |I_N|$ bits.

\subsection{DAC encoder}
\label{sec:dac_enc}

Similarly to other S-W coders, DAC is based on the principle of
inserting some ambiguity in the source description during the
encoding process. This is obtained using a modified interval
subdivision strategy. In particular, the DAC employs a set of
intervals whose lengths are proportional to the modified
probabilities $\widetilde{p}^X_{0}$ and $\widetilde{p}^X_{1}$, such
that $\widetilde{p}^X_{0} \geq p^X_{0}$ and $\widetilde{p}^X_{1}
\geq p^X_{1}$. In order to fit the enlarged sub-intervals into the
$[0,1)$ interval, they are allowed to partially overlap. This
prevents the decoder from discriminating the correct interval,
unless the side information is used.

\begin{figure}[!t]
\centering
\includegraphics[width=6cm,clip]{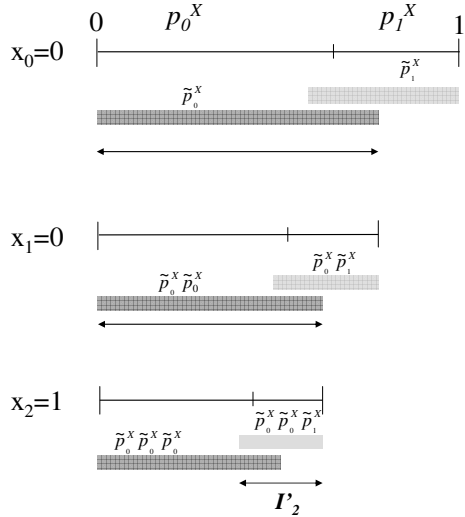}
\caption{Distributed arithmetic encoding procedure for a block of
three symbols.} \label{fig:dac}
\end{figure}

The detailed DAC encoding procedure is described in the following.
At initialization the ``current" interval is set to $I'_0=[0,1)$.
For each input symbol $x_i$, the current interval $I'_{i}$ is
subdivided into two partially overlapped sub-intervals whose lengths
are $\widetilde{p}^X_{0}|I'_{i}|$ and $\widetilde{p}^X_{1}|I'_{i}|$.
The interval representing symbol $x_i$ is selected as the next
current interval $I'_{i+1}$. After all $N$ symbols have been
processed, the sequence is represented by the final interval $I'_N$.
The codeword $C'_X$ can consist in the binary representation of any
number inside $I'_N$, and requires approximately $-\log_2 |I'_N|$
bits. This procedure is sketched in Fig.~\ref{fig:dac}. At the
decoder side, whenever the codeword points to an overlapped region,
the input symbol cannot be detected unambiguously, and additional
information must be exploited by the joint decoder to solve the
ambiguity. It is worth noticing that the DAC encoding procedure is a
generalization of AC. Letting $\widetilde{p}^X_{0}=p^X_{0}$ and
$\widetilde{p}^X_{1}=p^X_{1}$ leads to the AC encoding process
described in Sect. \ref{sec:ac}, with $I'_N=I_N$ and $C'_X=C_X$.

It should also be noted that, for simplicity, the description of the
AC and DAC provided above assumes infinite precision arithmetic. The
practical implementation used in Sect. \ref{sec:results_asymm} and
\ref{sec:results_symm} employs fixed-point arithmetic and interval
renormalization.

\section{Decoding for the asymmetric case}
\label{sec:dac_decoder}

The objective of the DAC decoder is joint decoding of the sequence
$\underline{x}$ given the correlated side information
$\underline{y}$. The arithmetic decoding machinery of the DAC
decoder presents limited modifications with respect to standard
arithmetic decoders; a fixed-point implementation has been employed,
with the same interval scaling and overlapping rules used at the
encoder. In the following the arithmetic decoder state at the $i$-th
decoding step is denoted as $\sigma_i, i=0,\ldots,N-1$. The data
stored in $\sigma_i$ represent the interval $I'_i$ and the codeword
at iteration $i$.

\begin{figure}[htb]
\centering
\includegraphics[width=10cm,clip]{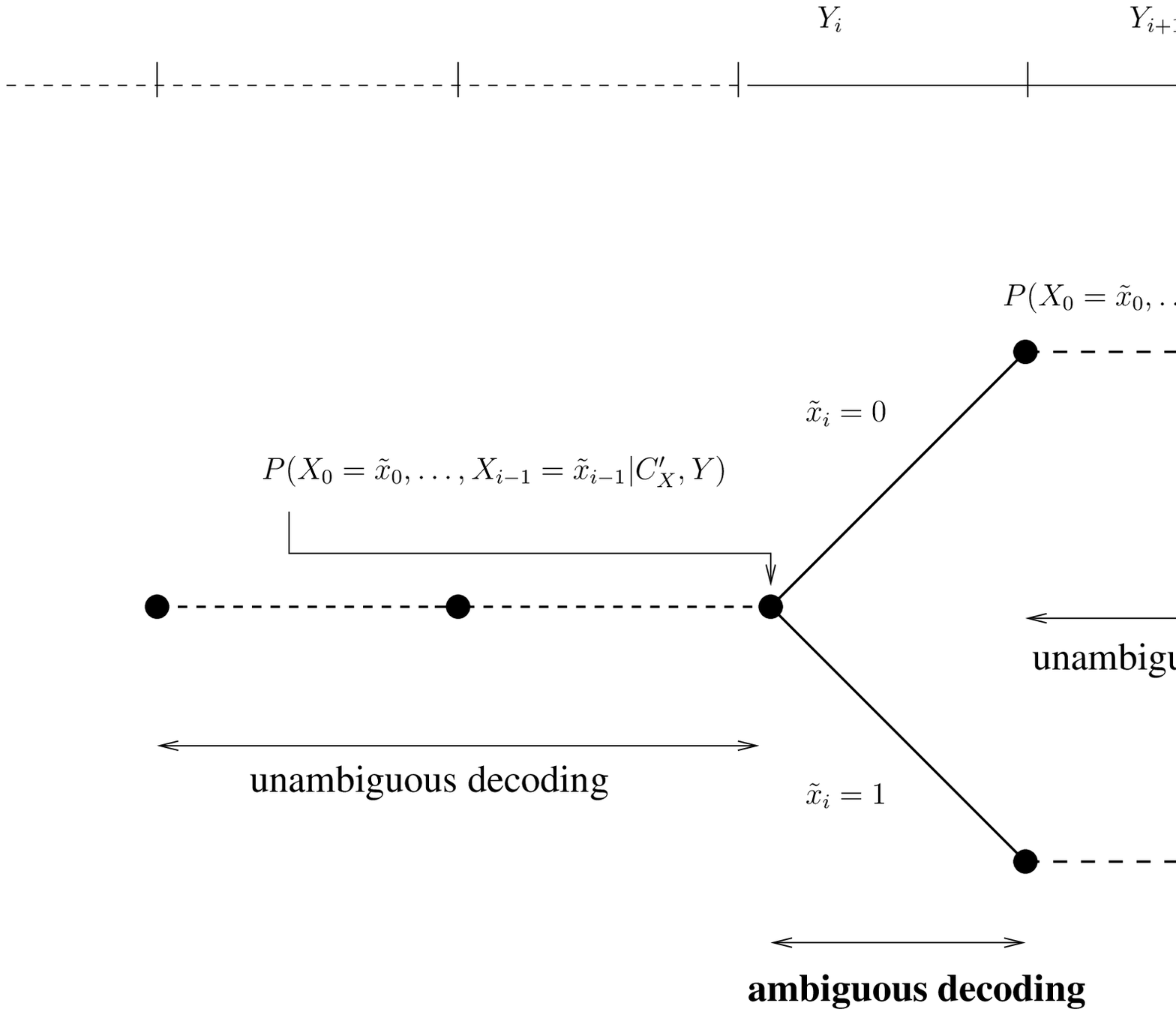}
\caption{Distributed arithmetic decoding tree for asymmetric S-W
coding.} \label{fig:dec_tree}
\end{figure}

The decoding process can be formulated as a symbol-driven sequential
search along a proper decoding tree, where each node represents a
state $\sigma_i$, and a path in the tree represents a possible
decoded sequence. The following elementary decoding functions are
required to explore the tree:
\begin{itemize}
 \item $(\tilde{x}_i,\sigma_{i+1})=${\it Test-One-Symbol}$(\sigma_i)$:
 it computes the sub-intervals at the $i$-th step, compares them with $C'_X$
 and outputs either an unambiguous symbol
    $\tilde{x}_i=0,1$ (if $C'_X$ belongs to one of the non-overlapped regions),
    or an ambiguous symbol $\tilde{x}_i=A$. In case of unambiguous decoding,
     the new decoder state $\sigma_{i+1}$ is returned for the following iterations.
\item $\sigma_{i+1}=${\it Force-One-Symbol}$(\sigma_i,\tilde{x}_i)$: it forces
the decoder to select the sub-interval corresponding to the symbol
$\tilde{x}_i$ regardless of the ambiguity; the updated decoder state
is returned.
\end{itemize}
In Fig.~\ref{fig:dec_tree} an example of a section of the decoding
tree is shown. In this example the decoder is not  able to make a
decision on the $i$-th symbol, as {\it Test-One-Symbol} returns
$\tilde{x}_i=A$. As a consequence, two alternative decoding attempts
are pursued by calling {\it Force-One-Symbol} with $\tilde{x}_i=0,1$
respectively. In principle, by iterating this process, the tree
$\mathcal{T}$, representing all the possible decoded sequences, can
be explored. The best decoded sequence can finally be selected
applying the {\em Maximum A Posteriori} (MAP) criterion
$\tilde{\underline{x}}= \arg\max_{\mathcal{T}}P(X_0,\ldots,X_{N-1} |
C'_X, Y)$.

In general, exhaustive search cannot be applied due to the
exponential growth of $\mathcal{T}$. A viable solution is obtained
applying the breadth-first sequential search known as $M$-algorithm
\cite{sa,tran_com05}; at each tree depth, only the $M$ nodes with
the best partial metric are retained. This amounts to visiting only
a subset of the most likely paths in $\mathcal{T}$. The MAP metric
for a given node can be evaluated as follows:
\begin{equation}
P(X_0=\tilde{x}_0,\ldots,X_{i}=\tilde{x}_i | C'_X, Y)=
\prod_{j=0}^{i} P(X_j=\tilde{x}_j|C'_X,Y_j) \label{eqn:linmet}
\end{equation}
Metric \eqref{eqn:linmet} can be expressed into additive terms by
setting:
\begin{eqnarray}
  \Lambda_{i+1} & \triangleq & \log  P(X_0=\tilde{x}_0,\ldots, X_i=\tilde{x}_i | C'_X , Y)  = \sum_{j=0}^i \lambda_j \\ \nonumber
  \lambda_j & \triangleq & \log  P(X_j=\tilde{x}_j|C'_X,Y_j)
\end{eqnarray}
\noindent where $\Lambda_0=0$ and $\lambda_i$  represent the
additive metric to be associated to each branch of $\mathcal{T}$.

The pseudocode for the DAC decoder is given in
Algorithm~\ref{alg:dac_decoder}, where $\mathcal{T}_i$ represents
the list of nodes in  $\mathcal{T}$ explored at depth $i$; each tree
node stores its corresponding arithmetic decoder state $\sigma_{i}$
and the accumulated metric $\Lambda_i$.

\begin{algorithm}[tb]
  \caption{DAC decoder (asymmetric case)}
  \label{alg:dac_decoder}
  \begin{algorithmic}
    \STATE  Initialize $\mathcal{T}_0$ with root node ($\sigma_0,\Lambda_0=0$)
    \STATE Set symbol counter $i \Leftarrow 0$
    \WHILE{ ($i < N$) }
        \FOR{ All nodes ($\sigma_i, \Lambda_i$) in $\mathcal{T}_i$ }
            \STATE $(\tilde{x}_i,\sigma_{i+1})=${\it Test-One-Symbol}$(\sigma_i)$
            \IF{$\tilde{x}_i=A$ }
            \FOR{$k=(0,1)$}
                \STATE $\sigma_{i+1}=${\it Force-One-Symbol}$(\sigma_i,\tilde{x}_i=k)$
                \STATE $\Lambda_{i+1} \Leftarrow \Lambda_i + \lambda_{i}$
                \STATE Insert $(\sigma_{i+1},\Lambda_{i+1})$ in $\mathcal{T}_{i+1}$
            \ENDFOR
            \ELSE
                \STATE $\Lambda_{i+1} \Leftarrow \Lambda_i + \lambda_i$
                \STATE Insert $(\sigma_{i+1},\Lambda_{i+1})$ in $\mathcal{T}_{i+1}$
            \ENDIF
        \ENDFOR
        \STATE Sort nodes in $\mathcal{T}_{i+1}$ according to metric $\Lambda_{i+1}$
        \STATE Keep only the $M$ nodes with best metric in $\mathcal{T}_{i+1}$
    \ENDWHILE
    \STATE Output $\tilde{\underline{x}}$ (sequence corresponding to the first node stored in $\mathcal{T}_{N}$)
  \end{algorithmic}
\end{algorithm}

It is worth pointing out that $M$ has to be selected as a trade-off
between the memory/complexity requirements and the error
probability, i.e., the probability that the path corresponding to
the original sequence $\underline{x}$ is accidentally dropped. As in
the case of standard Viterbi decoding, the path metric turns out to
be stable and reliable as long as a significant amount of terms,
i.e., number of decoded symbols $\tilde{x}_i$, are taken into
account. In the pessimistic case when all symbol positions $i$
trigger a decoder branching, given $M$, one can guarantee that at
least $\log_2(M)$ symbols are considered for metric comparisons and
pruning. On the other hand, in practical cases, the interval overlap
is only partial and branching does not occur at every symbol
iteration. All the experimental results presented in Sect.
\ref{sec:results_asymm} have been obtained using $M=2048$, while the
trade-off between performance and complexity is analyzed in Sect.
\ref{sec:perf_compl}.

Finally, metric reliability cannot be guaranteed for the very last
symbols of a finite-length sequence $\underline{x}$. For channel
codes, e.g., convolutional codes, this issue is tackled by imposing
a proper termination strategy, e.g., forcing the encoded sequence to
end in the first state of the trellis. A similar approach is
necessary when using DAC. Examples of AC termination strategies are
encoding a known termination pattern or end-of-block symbol with a
certain probability or, in the case of context-based AC, driving the
AC encoder in a given context. For DAC, we employ a new termination
policy that is tailored to its particular features. In particular,
termination is obtained by encoding the last $T$ symbols of the
sequence without interval overlap, i.e., using
$\widetilde{p}^X_{j}=p^X_{j}$, for all symbols $x_i$ with $i \geq
N-T$. As a consequence, no nodes in the DAC decoding tree will cause
branching in the last $T$ steps, making the final metrics more
reliable for the selection of the most likely sequence. However,
there is a rate penalty for the termination symbols.


\section{Rate allocation and choice of the overlap factor}
\label{sec:rate_sel}

The length of codeword $C'_X$ is determined by the length $|I'_N|$
of the final interval, which in turn depends on how much
$\widetilde{p}^X_{0}$ and $\widetilde{p}^X_{1}$ are larger than
$p^X_{0}$ and $p^X_{1}$. As a consequence, in order to select the
desired rate, it is important to quantitatively determine the
dependence of the expected rate on the overlap, because this will
drive the selection of the desired amount of overlap. Moreover, we
also need to understand how to split the overlap in order to achieve
good decoding performance. In the following we derive the expected
rate obtained by the DAC as a function of the set of input
probabilities and the amount of overlap.

\subsection{Calculation of the rate yielded by DAC}

We are interested in finding the expected rate $\widetilde{R}$ (in
bps) of the codeword used by the DAC to encode the sequence
$\underline{x}$. This is given by the following formula:
%
\begin{equation} \label{eq:rate1}
\widetilde{R}=\sum_{j=0}^1 p^X_{j} \log_2
\frac{1}{\widetilde{p}^X_{j}}
\end{equation}
\noindent This can be derived straightforwardly from the property
that the codeword generated by an AC has an expected length that
depends on the size of the final interval, that is, on the product
of the probabilities $\widetilde{p}^X_{j}$, and hence on the amount
of overlap. The expectation is computed using the true probabilities
$p^X_{j}$.

We set $\widetilde{p}^X_{j}=\alpha^X_{j} p^X_{j}$, where
$\alpha^X_{j} \geq 1$, so that
$\widetilde{p}^X_{0}+\widetilde{p}^X_{1} \geq 1$. This amounts to
enlarging each interval by an amount proportional to the overlap
factors $\alpha^X_{j}$. The expected rate achieved by the DAC
becomes
\[
\widetilde{R}= \sum_{j=0}^1 p^X_{j} \left( r^X_{j} - \delta^X_{j}
\right)
\]
\noindent where $r^X_{j}=-\log_2 p^X_{j}$, and $\delta^X_{j}=\log_2
\alpha^X_{j}$. Note that $r^X_{j}$ represents the rate contribution
of symbol $j$ yielded by standard AC, while $\delta^X_{j}$
represents the decrease of this contribution, i.e., the average
number of bits saved in the binary representation of the $j$-th
input symbol.

\subsection{Design of the overlap factors}

Once a target rate has been selected, the problem arises of
selecting  $\alpha ^X_{j}$.
As an example, a possible choice is to take equal overlap factors
$\alpha ^X_{0}=\alpha ^X_{1}=\alpha ^X$. This implies that each
interval is enlarged by a factor $\alpha ^X$ that does not depend on
the source probability $p^X_j$. This leads to a target rate
\begin{equation}
\label{eq:alpha_const} R'_X=H(X) - \log_2 \alpha ^X.
\end{equation}
It can be shown that this choice minimizes the rate $\widetilde{R}$
for a given total amount of overlap $\alpha ^X_{0}p^X_0 + \alpha
^X_{1}p^X_1-1$; the computations are simple and are omitted for
brevity. This choice is not necessarily optimal in terms of the
decoder error probability. However, optimizing for the error
probability is impractical because of the nonlinearity of the
arithmetic coding process.

In practice, one also has to make sure that the enlarged intervals
$[0, \alpha ^X_{0}p^X_0)$ and $[1- \alpha ^X_{1} p^X_1,1)$ are both
contained inside the $[0,1)$ interval. E.g., taking equal overlap
factors as above does not guarantee this. We have devised the
following rule that allows to achieve any desired rate satisfying
the constraint above. We apply the following constraint:
\begin{equation} \label{eq:kappa}
\frac{\delta^X_{j}}{r^X_{j}} = k^X
\end{equation}
\noindent with $k^X$ a positive constant independent of $j$. This
leads to
\begin{equation} \label{eq:criterion}
\alpha^X_{j}=(p^X_{j})^{-k^X}
\end{equation}
This can be interpreted as an additional constraint that the rate
reduction for symbols ``0" and ``1" depends on their probabilities,
i.e., the least probable symbol undergoes a larger reduction. Using
(\ref{eq:criterion}), it can be easily shown that the expected rate
achieved by the DAC can be written as
\begin{equation}
\label{eq:3} \widetilde{R}= \left( 1-k^X \right) H(X).
\end{equation}

Thus, the allocation problem for an i.i.d. source is very simple. We
assume that the conditional entropy $H(X|Y)$ is available as in Fig.
\ref{fig:modeling}-b, modeling the correlation between $X$ and $Y$.
In asymmetric DSC, $\underline{x}$ should be ideally coded at a rate
arbitrarily close to $H(X|Y)$. In practice, due to the suboptimality
of any practical coder, some margin $\mu \geq 1$ should be taken.
Hence, we assume that the allocation problem can be written as
$\left( 1-k^X \right) H(X)\leq \mu H(X | Y) $. Since $\mu$ is a
constant and $H(X | Y)$ and $H(X)$ are given, one can solve for
$k^X$ and then perform the encoding process.

Finally, it should be noted that, while we have assumed that $X$ and
$Y$ are i.i.d., the DAC concept can be easily extended to a
nonstationary source. This simply requires to consider all
probabilities and overlap factors as depending on index $i$; all
computations, including the design of the overlap factors and the
derivation of the target rate, can be extended straightforwardly. A
possible application is represented by context-based coding or
Markov modeling of correlated sources. There is one caveat though,
in that, if the probabilities and context of each symbol are
computed by the decoder from past symbols, decoding errors can
generate significant error propagation.

\section{Distributed arithmetic coding: the symmetric case}
\label{sec:symm}


\subsection{Symmetric DAC encoding and rate allocation}

In many applications, it is preferable to encode the correlated
sources at similar rather than unbalanced rates; in this case,
symmetric S-W coding can be used. Considering a pair of sources, in
symmetric S-W coding both $X$ and $Y$ are encoded using separate
DACs. We denote as $C'_X$ and $C'_Y$ the codewords representing $X$
and $Y$, and $R'_X$ and $R'_Y$ the respective rates. With DAC, the
rate of $X$ and $Y$ can be adjusted with a proper selection of the
parameters $k^X$ and $k^Y$ for the two DAC encoders. However, it
should be noted that, for the same total rate, not all possible
choices of $k^X$ and $k^Y$ are equally good, because some of them
could complicate the decoder design, or be suboptimal in terms of
error probability. To highlight the potential problems of a
straightforward extension of the asymmetric DAC, let us assume that
$k^X$ and $k^Y$ can be chosen arbitrarily. This would require a
decoder that performs a search in a symbol-synchronous tree where
each node represents {\em two} sequential decoder states
$(\sigma^X_i, \sigma^Y_i)$ for $X$ and $Y$ respectively. If the
interval selection is ambiguous for both sequences, the four
possible binary symbol pairs (00,01,10,11) need to be included in
the search space; this would accelerate the exponential growth of
the tree, and quickly make the decoder search unfeasible. This
example shows that some constraints need to be put on $k^X$ and
$k^Y$ in order to limit the growth rate of the search space.

To overcome this problem, we propose an algorithm that applies the
idea of time-sharing to the DAC. The concept of time-shared DAC has
been preliminarly presented in \cite{tsdac} for a pair of sources in
the subcase $R'_X=R'_Y$, i.e. providing only the mid-point of the
S-W rate region. In the following we extend this to an arbitrary
combination of rates, and show how this can be generalized to an
arbitrary number of sources. For two sources, the idea is to divide
the set of input indexes $i=0, 1, \ldots , N-1$ in two disjoint sets
such that, at each index $i$, ambiguity is introduced in at most one
out of the two sources. In particular, for sequences $\underline{x}$
and $\underline{y}$ of length $N$, let ${\cal A}_X$ and ${\cal A}_Y$
be the subsets of even and odd integer numbers in $\{ 0, \ldots,
N-1\}$ respectively. We employ a DAC on $\underline{x}$ and
$\underline{y}$, but the choice of parameters $k^X$ and $k^Y$
differs. In particular, we let the parameters depend on the symbol
index $i$, i.e., $k^X_i$ and $k^Y_i$. The DAC of $\underline{x}$
employs parameter $k^X_i=k^X \geq 0$ for all $i \in {\cal A}_X$, and
$k^X_{i}=0$ otherwise. Vice versa, $\underline{y}$ is encoded with
parameter $k^Y_{i}=k^Y\geq 0$ for all $i \in {\cal A}_Y$, and
$k^Y_{i}=0$ otherwise. As a consequence of these constraints, at
each step of the decoding process, ambiguity appears in at most one
out the two sequences. In this way, the growth rate of the decoding
tree remains manageable, as no more than two new states are
generated at each transition, exactly as in the asymmetric DAC
decoder; this also makes the MAP metric simpler. The conceptual
relation with time-sharing is evident. Since, during the DAC
encoding process, for each input symbol the ambiguity is introduced
in at most one out the two encoders, this corresponds to switching
the role of side information between either source on a
symbol-by-symbol basis.

By varying the parameters $k^X$ and $k^Y$, all combinations of rates
can be achieved. The achieved rates can be derived repeating the
same computations described in Sect. \ref{sec:rate_sel}, and can be
expressed as $R'_X = \left( 1 - \frac{k^X}{2}  \right) H(X)$ and
$R'_Y = \left( 1 - \frac{k^Y}{2} \right) H(Y)$. The rate allocation
problem amounts to selecting suitable rates $R'_X$ and $R'_Y$ such
that $R'_X \geq H(X|Y)$, $R'_Y \geq H(Y|X)$, and $R'_X+R'_Y \geq
H(X,Y)$. In practice one will typically take some margin $\mu \geq
1$, such that $R'_X+R'_Y = \mu H(X,Y)$; for safety, a margin should
also be taken on $R'_X$ and $R'_Y$ with respect to the conditional
entropy. Since the prior probabilities of $X$ and $Y$ are given, one
can solve for $k^X$ and $k^Y$, and then perform the encoding
process. Thus, the whole S-W rate region can be swept.

\subsection{Decoding process for symmetric DAC}

Similarly to the asymmetric case, the symmetric decoding process can
be viewed as a search along a tree; however, specifically for the
case of two correlated sources, each node in tree represents the
decoding states $(\sigma_i^X,\sigma_i^Y)$ of two sequential
arithmetic decoders for $\underline{x}$ and $\underline{y}$
respectively. At each iteration, sequential decoding is run from
both states. The time-sharing approach guarantees that, for a given
index $i$, the ambiguity can be found only in one of the two
decoders. Therefore, at most two branches must be considered, and
the tree can be constructed using the same functions introduced in
Sect. \ref{sec:dac_decoder} for the asymmetric case. This would be
the same also for $P$ sources. In particular, for $i \in
\mathcal{A}_X$, {\em Test-One-Symbol}($\sigma_i^Y$) yields an
unambiguous symbol $\tilde{y}_i \neq A$, whereas ambiguity can be
found only while attempting decoding for $\underline{x}$ with {\em
Test-One-Symbol}($\sigma_i^Y$). In conclusion, from the node
$(\sigma_i^X,\sigma_i^Y)$ the function {\em Test-One-Symbol} is used
on both states. If ambiguity is found on $\tilde{x}_i$, {\em
Force-One-Symbol} is then used to explore the two alternative paths
for $\tilde{x}_i$, whereas $\tilde{y}_i$ is used as side information
for branch metric evaluation. In the case that $i\in {\cal A}_Y$,
the roles of $\underline{x}$ and $\underline{y}$ are exchanged.
Therefore, Algorithm \ref{alg:dac_decoder} can be easily extended to
the symmetric case by alternatively probing either $\underline{x}$
or $\underline{y}$ for ambiguity, and possibly generating a
branching. The joint probability distribution can be written as
\begin{equation}\label{eqn:metric}
P(X_0=\tilde{x}_0,
\ldots,X_{N-1}=\tilde{x}_{N-1},Y_0=\tilde{y}_0,\ldots,Y_{N-1}=\tilde{y}_{N-1}
| C'_X,C'_Y ) =
\end{equation}
\[
=\prod_{i \in {\cal A}_X } P(X_i=\tilde{x}_i | Y_i, C'_X,C'_Y)
\prod_{i \in {\cal A}_Y } P(Y_i=\tilde{y}_i|X_i, C'_X,C'_Y)
\]

The symmetric encoder and decoder can be easily generalized to an
arbitrary number $P$ of sources. The idea is to identify $P$ subsets
of input indexes $i=0, 1, \ldots , N-1$ such that, at each symbol
index $i$, ambiguity is introduced in at most one out of the $P$
sources. In particular, for sequences $\underline{x}^{(1)}, \ldots,
\underline{x}^{(P)}$ of length $N$, let ${\cal A}_1, \ldots, {\cal
A}_P$ be disjoint subsets of $\{ 0, 1, \ldots, N-1\}$. We denote the
DAC parameters as $k^{(1)}_i, \ldots, k^{(P)}_i$. The DAC of
$\underline{x}^{(j)}$ employs parameter $k^{(j)}_i=k^{(j)} \geq 0$
for all $i \in {\cal A}_j$, and $k^{(j)}_{i}=0$ otherwise. As a
consequence of these constraints, at each step of the decoding
process, ambiguity appears in at most one out the $P$ sequences.
Note that this formulation also encompasses the case that one or
more sources are independent of each other and from all the others;
these sources can be coded with a classical AC, taking ${\cal A}_j =
\emptyset$ for this source.

The selection of the sets ${\cal A}_j$ and the overlap factors
$k^{(j)}$, for $j=1, \ldots,P$, is still somewhat arbitrary, as the
expected rate of source $j$ depends on both the cardinality of
${\cal A}_j$ and the value of $k^{(j)}$. In a realistic application
it would be more practical to fix the sets ${\cal A}_j$ once and for
all, and to modify the parameters $k^{(j)}$ so as to obtain the
desired rate. This is because, for time-varying correlations, one
has to update the rate on-the-fly. In a distributed setting, varying
one parameter $k^{(j)}$ requires to communicate the change only to
source $j$, while varying the sets ${\cal A}_j$ requires to
communicate the change to all sources. Therefore, we define ${\cal
A}_j$ such that the $P$ statistically dependent sources take in
turns the role of the side information. Any additional independent
sources are coded separately using ${\cal A}_j = \emptyset$. In
particular, we set ${\cal A}_j = \{ k| k\%P =j \}$, where $\%$
denotes the remainder of the division between two integers, and
$0\%j=0$. The DAC encoder for the $j$-th source inserts ambiguity
only at time instants $i \in {\cal A}_j$. At each node, the decoder
stores the states of the $P$ arithmetic decoders, and possibly
performs a branching if the codeword related to the only potentially
ambiguous symbol at the current time $i$ is actually ambiguous.
Although this encoding and decoding structure is not necessarily
optimal, it does lead to a viable decoding strategy.

\section{Results: asymmetric coding}
\label{sec:results_asymm}

In the following we provide results of a performance evaluation
carried out on DAC. We implement a communication system that employs
a DAC and a joint decoder, with no feed-back channel; at the
decoder, pruning is performed using the M-algorithm \cite{sa}, with
M=2048. The side information is obtained by sending the source $X$
through a binary symmetric channel with transition probability $p$,
which measures the correlation between the two sources. We simulate
a source with both balanced ($p_0=0.5$) and skewed ($p_0 > 0.5$)
symbol probabilities. The first setting implies $H(X)=H(Y)=1$ and
$H(X,Y)=1+H(X|Y)$, where $H(X|Y)$ depends on $p$. The closer $p$ to
0.5, the less correlated the sources, and hence the higher $H(X|Y)$.
In the skewed case, given $p_0$, $H(X)$ is fixed, whereas both
$H(Y)$ and $H(X|Y)$ depend on $p$. Unless otherwise specified, each
point of the figures/tables presented in the following has been
generated averaging the results obtained encoding $10^7$ samples.

\subsection{Effect of termination}

As a first experiment, the benefit of the termination policy is
assessed. An i.i.d. stationary source $X$ emits sequences
$\underline{x}$ of $N=200$ symbols, with $p_0=0.5$ and
$H(X|Y)=0.25$, which are encoded with DAC at fixed rate $0.5$ bps,
i.e., $0.25$ bps higher than the theoretical S-W bound. For $Y$ we
assume ideal lossless encoding at average rate $H(Y)=1$ bps, so that
the total average rate of $X$ and $Y$ is 1.5 bps. The bit error rate
(BER) yielded by the decoder is measured for increasing values of
the number of termination symbols $T$. The same simulation is
performed with $N=1000$. In all simulated cases, the DAC overlap has
been selected to compensate for the rate penalty incurred by the
termination, so as to achieve the 1.5 bps overall target rate. The
overlap factors $\alpha_{j}^X$ are selected according to
\eqref{eq:criterion}.

\begin{figure}[htb]
\centering
\includegraphics[width=12cm,clip]{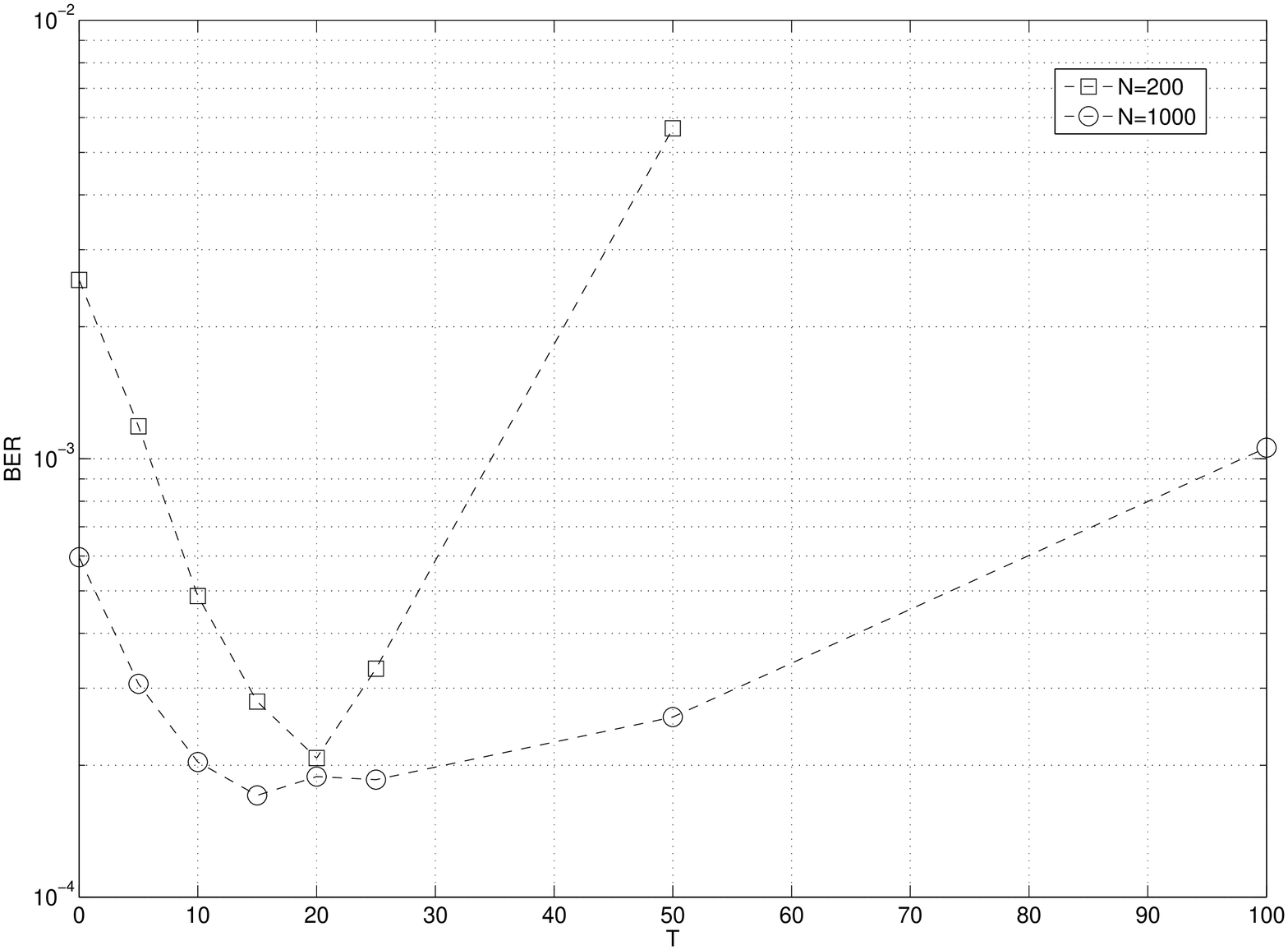}
\caption{BER as function of  $T$ (number of termination symbols);
$p_0=0.5$, total rate = 1.5 bps, rate of $\underline{x}$ = 0.5 bps,
$H(X|Y)=0.25$.} \label{fig:term}
\end{figure}

The results are shown in Fig.~\ref{fig:term}; it can be seen that
the proposed termination is effective at reducing the BER. There is
a trade-off in that, for a given rate, increasing $T$ reduces the
effect of errors in the last symbols, but requires to overlap the
intervals more. It is also interesting to consider the position of
the first decoding error as, without termination, errors tend to
cluster at the end of the block. For $N=200$, the mean position
value is 191, 178, 168, 161 and 95, with standard deviation 13, 18,
25, 36 and 49, respectively for  $T$ equal to 0, 5, 10, 15 and 20.
For $N=1000$, the mean value is 987, 954, 881, 637 and 536, with
standard deviation 57, 124, 229, 308 and 299. The optimal values of
$T$ are around 15-20 symbols. Therefore, we have selected $T=15$ and
used this value for all the experiments reported in the following.

\subsection{Effect of the overlap design rule}

\begin{figure}[htb]
\centering
\includegraphics[width=12cm,clip]{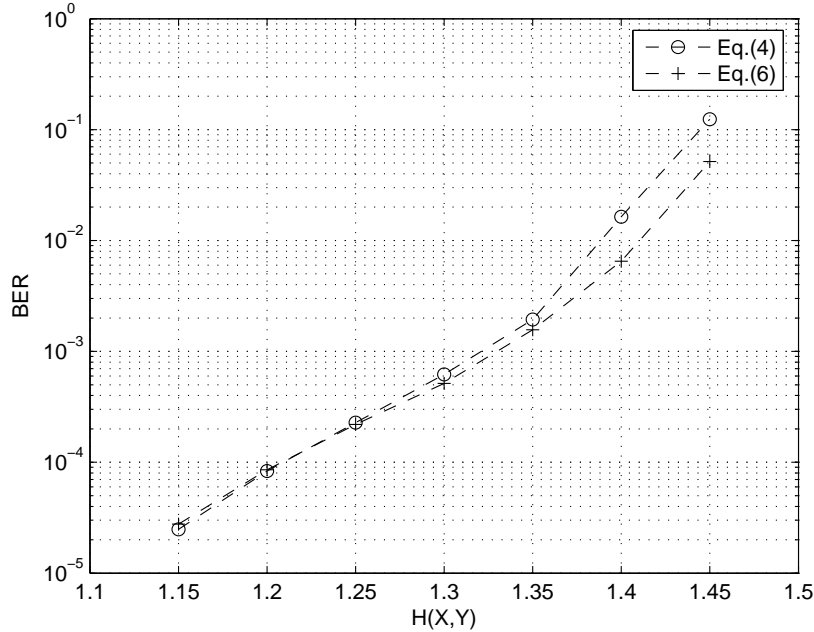}
\caption{Performance comparison between the use of different overlap rules ($p_0=0.8$, total rate = 1.5 bps).} \label{fig:overlap}
\end{figure}

Next, an experiment has been performed to validate the theoretical
analysis of the effects of different overlap designs shown in
Sect.~\ref{sec:rate_sel}-B. In Fig. \ref{fig:overlap} the
performance obtained by using the design of equations
\eqref{eq:alpha_const} and \eqref{eq:criterion} respectively is
shown. The experimental settings are $N=200$, $p_0=0.8$, fixed rate
for $\underline{x}$ of 0.5 bps, and total average rate for $X$ and
$Y$ equal to 1.5 bps, with ideal lossless encoding of $Y$ at rate
$H(Y)$.
The BER is reported as a function of the source correlation
expressed in terms of $H(X,Y)$. It is worth noticing that the
performance yielded by different overlap design rules are almost
equivalent. Note that the rule in \eqref{eq:criterion} consistently
outperforms that in \eqref{eq:alpha_const}, confirming that this
latter is only optimal for the rate. There is some difference when
$H(X,Y)$ is very high (i.e., for weakly correlated sources).
However, this case is of marginal interest since the performance is
poor (the BER is of the order of 0.1).

\subsection{Performance evaluation at fixed rate}

The performance of the proposed system is compared with that of a
system where the DAC encoder and decoder are replaced by a punctured
turbo code similar to that in \cite{aaron}. We use turbo codes with
rate-$\frac{1}{2}$ generator (17,15) octal (8 states) and (31,27)
octal (16 states), and employ S-random interleavers, and 15 decoder
iterations. We consider the case of balanced source ($p_0=p_1=0.5$)
and skewed source (in particular $p_0=0.9$ and $p_0=0.8$). For a
skewed source, as an improvement with respect to \cite{aaron}, the
turbo decoder has been modified by adding to the decoder metric the
{\em a priori} term, as done in \cite{frias}. Block sizes $N=50$,
$N=200$ and $N=1000$ have been considered (with S-random interleaver
spread of 5, 11 and 25 respectively); this allows to assess the DAC
performance at small and medium block lengths. Besides turbo codes,
we also considered the rate-compatible LDPC codes proposed in
\cite{varodayan}. For these codes, a software implementation is
publicly available on the web; among the available pre-designed
codes, we used the matrix for $N=396$, which is comparable with the
block lengths considered for the DAC and the turbo code.

\begin{figure}[htb]
\centering
\includegraphics[width=12cm,clip]{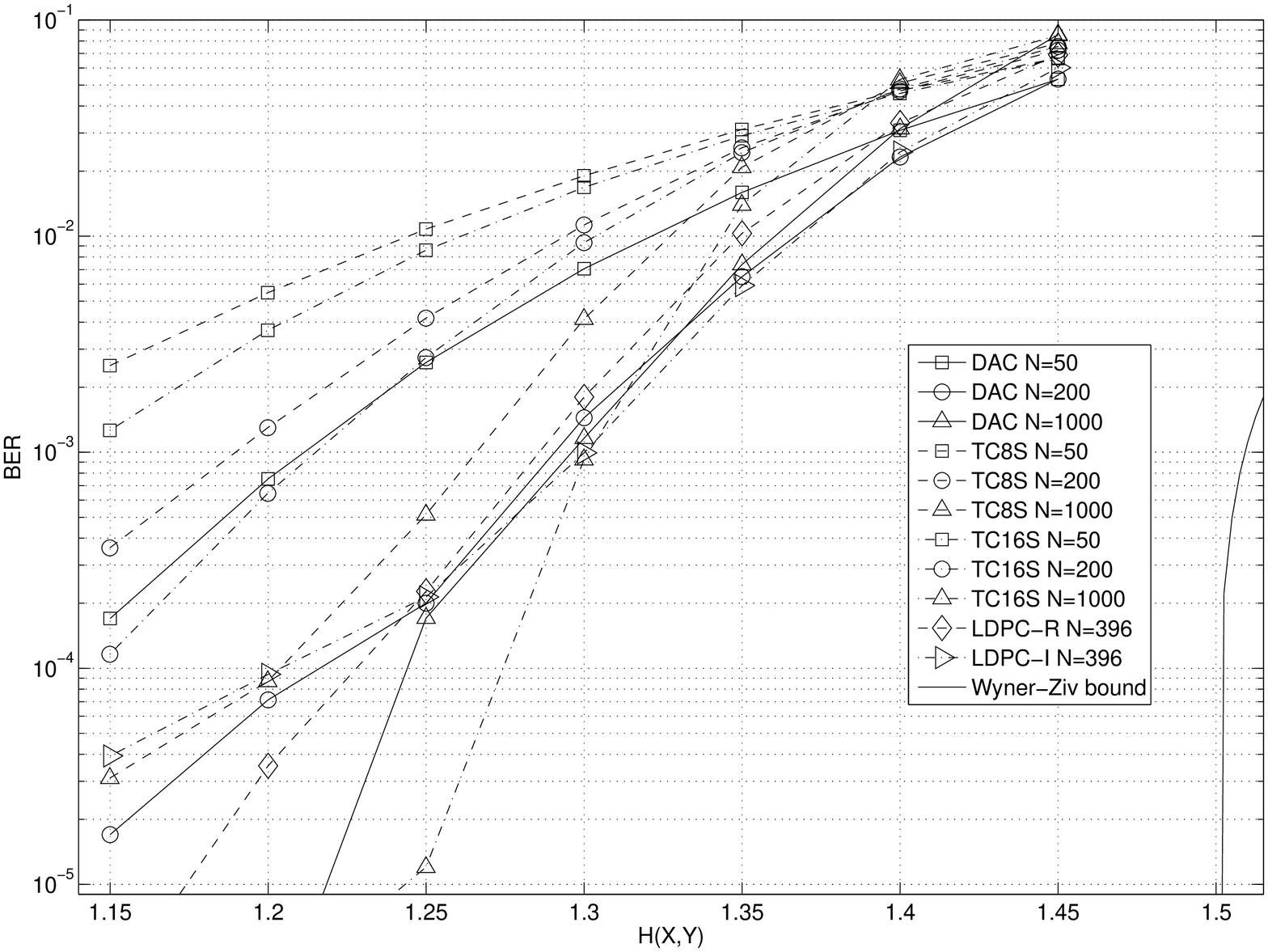}
\caption{Performance comparison of data communication systems
($p_0=0.5$, total rate = 1.5 bps, rate for $\underline{x}$ = 0.5
bps): DAC versus turbo coding, balanced source. DAC: distributed
arithmetic coding; TC8S and TC16S: 8- and 16-state turbo code with
S-random interleaver; LDPC-R and LDPC-I: regular and irregular LDPC
codes from \cite{varodayan}.} \label{fig:1}
\end{figure}

The results are worked out in a fixed-rate coding setting as in
\cite{xiong_spm}, i.e., the rate is the same for each sample
realization of the source. Fig. \ref{fig:1} reports the results for
the balanced source case; the abscissa is $H(X,Y)$, and is related
to $p$.
The performance is measured in terms of the residual BER after
decoding, which is akin to the distortion in the Wyner-Ziv binary
coding problem with Hamming metric. Both the DAC and the turbo code
generate a description of $\underline{x}$ at fixed rate 0.5 bps; the
total average rate of $X$ and $Y$ is 1.5 bps, with ideal lossless
encoding of $Y$ at rate $H(Y)$. Since $H(Y)=1$, we also have that
$H(X,Y)=1+H(X|Y)$. This makes it possible to compare these results
with the case of skewed sources which is presented later in this
section, so as to verify that the performance is uniformly good for
all distributions. The Wyner-Ziv bound for a doubly symmetric binary
source with Hamming metric is also reported for comparison.

As can be seen, the performance of DAC slightly improves as the block length
increases. This is mostly due to the effect of the termination. As the number
of bits used to terminate the encoder is chosen independently of the block
length, the rate penalty for non overlapping the last bits weights more when
the block length is small, while the effect vanishes for large block length. In
\cite{dac_comlet}, where the termination effect is not considered, the
performance is shown to be almost independent of the block size. It should also
be noted that the value of $M$ required for near-optimal performance grows
exponentially with the block size. As a consequence, the memory which leads to
near-optimal performance for $N=50$ or $N=200$ limits the performance for
$N=1000$.

We compared both 8-states and 16-states turbo codes. The 8-states
code is often used in practical applications, as it exhibits a good
trade-off between performance and complexity; the 16-states code is
more powerful, and requires more computations. It can be seen that,
for block length $N=50$ and $N=200$, the proposed system outperforms
the 8-states and 16-states turbo codes. For block length $N=1000$,
the DAC performs better than the 8-states turbo code, and is
equivalent to the 16-states code. It should be noted that, in this
experiment, only the ``channel coding performance" of the DAC is
tested, since for the balanced source no compression is possible as
$H(X)=1$. Consequently, it is remarkable that the DAC turns out to
be generally more powerful than the turbo code at equal block
length. Note that the performance of the 16-states code is limited
by the error floor, and could be improved using an ad-hoc design of
the code or the interleaver; the DAC has no error floor, but its
waterfall is less steep. For $H(X|Y)\geq 0.35$, a result not
reported in Fig. \ref{fig:1} shows that the DAC with $N=200$ and
$N=1000$ also outperform the 8-state turbo-coder with $N=5000$. In
Fig. \ref{fig:1} and in the following, it can be seen that turbo
codes do not show the typical cliff-effect. This is due to the fact
that, at the block lengths considered in this paper, the turbo code
is still very far from the capacity; its performance improves for
larger block lengths, where the cliff-effect can be seen.
In terms of the rate penalty, setting a residual BER
threshold of $10^{-4}$, for $N=200$ the DAC is almost 0.3 bps away
from the S-W limit, while the best 16-state turbo code simulated in
this paper is 0.35 bps away; for $N=1000$  the DAC is 0.26 bpp away,
while the best 8-state turbo code is 0.30 bps away. The performance
of the LDPC code for $N=396$ is halfway between the turbo codes for
$N=200$ and $N=1000$, and hence very similar to the DAC.

\begin{figure}[htb]
\centering
\includegraphics[width=12cm,clip]{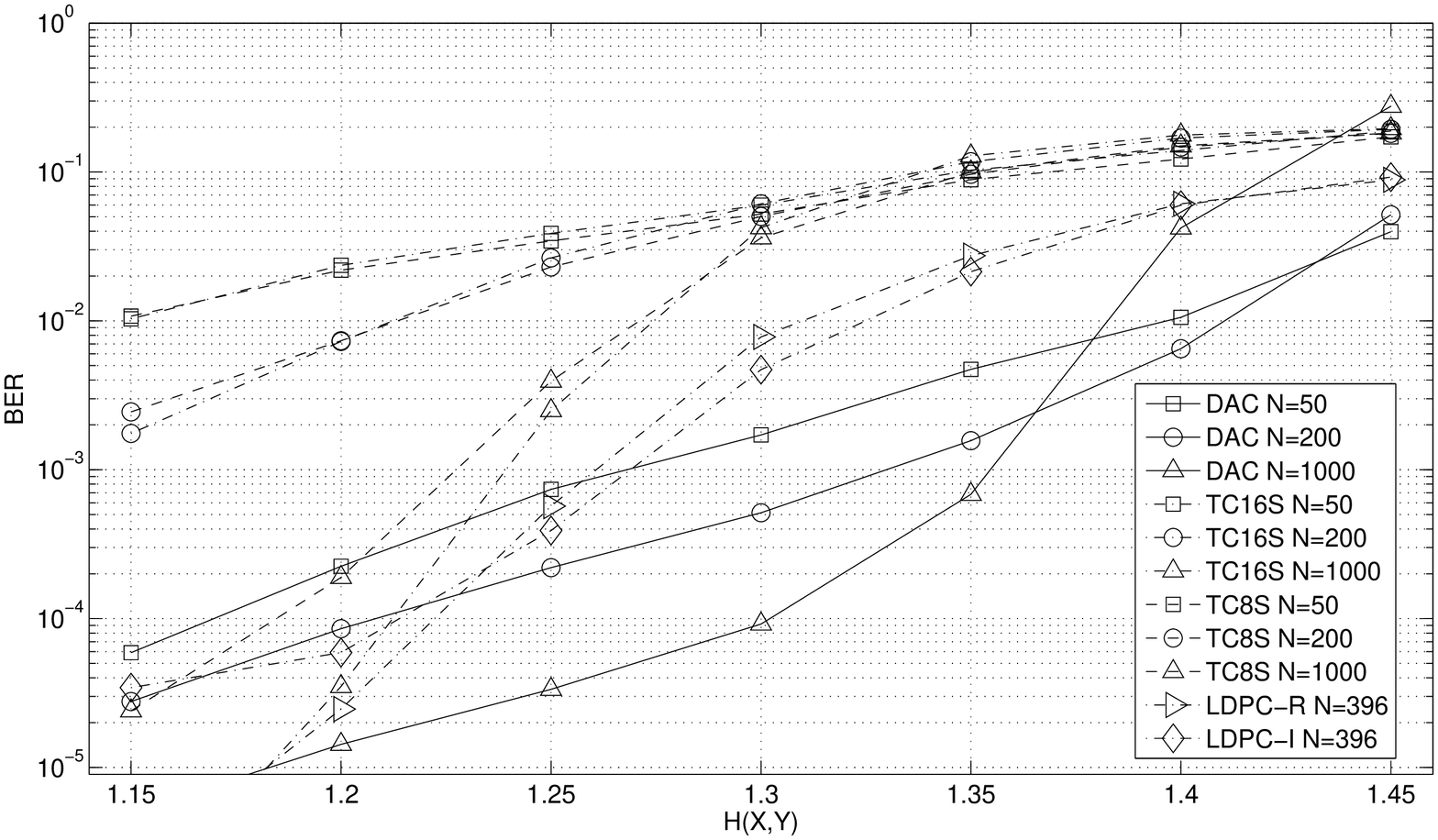}
\caption{Performance comparison of data communication systems
($p_0=0.8$, total rate = 1.5 bps): DAC versus turbo coding, skewed
source. DAC: distributed arithmetic coding; TC8S and TC16S: 8- and
16-state turbo code with S-random interleaver; LDPC-R and LDPC-I:
regular and irregular LDPC codes from \cite{varodayan}.}
\label{fig:3}
\end{figure}

The results for a skewed source are reported in Fig. \ref{fig:3} for
$p_0=0.8$. In this setting, we select various values of $H(X,Y)$,
and encode $\underline{x}$ at fixed rate such that the total average
rate for $X$ and $Y$ equals 1.5 bps, with ideal lossless
encoding of $Y$ at rate $H(Y)$. 
For Fig. \ref{fig:3}, from left to right, the rates of
$\underline{x}$ are respectively 0.68, 0.67, 0.66, 0.64, 0.63, 0.61,
0.59, and 0.58 bps.
Consistently with \cite{tiffany}, all turbo codes considered in this
work perform rather poorly on skewed sources. In \cite{tiffany} this
behavior is explained with the fact that, when the source is skewed,
the states of the turbo code are used with uneven probability,
leading to a smaller equivalent number of states. On the other hand,
the DAC has good performance also for skewed sources, as it is
designed to work with unbalanced distributions. The performance of
the LDPC codes is similar to that of the best turbo codes, and
slightly worse than the DAC.

\begin{figure}[htb]
\centering
\includegraphics[width=12cm,clip]{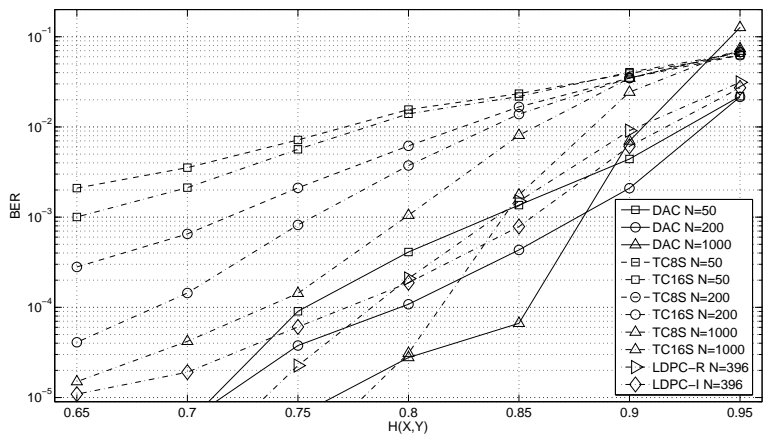}
\caption{Performance comparison of data communication systems
($p_0=0.9$, total rate = 1 bps): DAC versus turbo coding, skewed
source. DAC: distributed arithmetic coding; TC8S and TC16S: 8- and
16-state turbo code with S-random interleaver; LDPC-R and LDPC-I:
regular and irregular LDPC codes from \cite{varodayan}.}
\label{fig:2}
\end{figure}

Similar remarks can be made in the case of $p_0=0.9$, which is
reported in Fig. \ref{fig:2}. In this case, we have selected a total
rate of 1 bps, since the source is more unbalanced and hence easier
to compress. The rates for $\underline{x}$ are respectively 0.31,
0.34, 0.37, 0.39, 0.42, 0.44, and 0.47 bps. In this case the turbo
code performance is better than in the previous case, although it is
still poorer than DAC. This is due to the fact that the sources are
more correlated, and hence the crossover probability on the virtual
channel is lower. Therefore, the turbo code has to correct a smaller
number of errors, whereas for $p_0=0.8$ the correlation was weaker
and hence the crossover probability was higher.


\subsection{Performance evaluation for strongly correlated sources}

We also considered the case of strongly correlated sources, for
which high-rate channel codes are needed. These sources are a good
model for the most significant bit-planes of several multimedia
signals. Due to the inefficiency of syndrome-based coders, practical
schemes often assume that no DSC is carried out on those bit-planes,
e.g., they are not transmitted, and at the decoder they are directly
replaced by the side information \cite{xiong_tip}.

The results are reported in Tab. \ref{tab:corr} for the DAC and the
16-state turbo code, when a rate of 0.1 bps is used for
$\underline{x}$. The table also reports the cross-over probability
$p$, corresponding, for a balanced source, to the performance of an
uncoded system that reconstructs $\underline{x}$ as the side
information $\underline{y}$. As can be seen, the DAC has similar
performance to the turbo codes and LDPC codes, and becomes better
when the source is extremely correlated, i.e., $H(X|Y)=0.001$.

\begin{table}[htb]
\caption{Residual BER in case of strongly correlated sources, with
 $p_0=0.5$ and rate for $\underline{x}$ equal to 0.1 bps.}
\begin{center}
\begin{tabular}{|c|c||c|c|}  \hline
\multicolumn{4}{|c|}{$N=200$} \\
\hline
$H(X|Y)$ & $p$ &  DAC  & TC16S \\
 \hline
0.1 & $1.3\cdot 10^{-2}$ & $2.25\cdot 10^{-2}$ &  $1.05\cdot 10^{-2}$ \\
0.01 & $8.6\cdot 10^{-4}$ & $2.55\cdot 10^{-4}$  & $1.74\cdot 10^{-4}$ \\
0.001 & $6.5\cdot 10^{-5}$ & $1.5\cdot 10^{-6}$  & $7.0\cdot 10^{-6}$ \\
\hline

\multicolumn{4}{|c|}{$N=1000$} \\
\hline
$H(X|Y)$ &  $p$ & DAC & TC16S \\
 \hline
0.1 & $1.3\cdot 10^{-2}$ & $2.10\cdot 10^{-2}$ & $1.18\cdot 10^{-2}$ \\
0.01  & $8.6\cdot 10^{-4}$ & $1.5\cdot 10^{-5}$ & $2.9\cdot 10^{-5}$ \\
0.001   & $6.5\cdot 10^{-5}$ & $<  1 \cdot 10^{-6}$ & $1.0\cdot 10^{-6}$ \\
\hline

\multicolumn{4}{|c|}{$N=396$} \\
\hline
$H(X|Y)$ & $p$ & LDPC-R & LDPC-I \\
 \hline
0.1 & $1.3\cdot 10^{-2}$ & $1.20\cdot 10^{-2}$ & $1.11\cdot 10^{-2}$ \\
0.01  & $8.6\cdot 10^{-4}$ & $1.18\cdot 10^{-4}$ & $1.01\cdot 10^{-4}$ \\
0.001   & $6.5\cdot 10^{-5}$ & $4.65\cdot 10^{-6}$ & $7.58\cdot 10^{-6}$ \\
\hline

\end{tabular}
\end{center}
\label{tab:corr}
\end{table}

\subsection{Performance evaluation at variable rate}
\label{sec:res_vr}

Finally, the coding efficiency of DAC is measured in terms of
expected rate required to achieve error-free decoding. This amounts
to re-encoding the sequence at increasing rates, and represents the
optimal DAC performance if the encoder could exactly predict the
decoder behavior. Since each realization of the source is encoded
using a different number of bits, this case is referred to as
variable-rate encoding. This scenario is representative of practical
distributed compression settings, e.g., \cite{aaron}, in which one
seeks the shortest code that allows to reconstruct without errors
each realization of the source process.

For this simulation, the following setup is used. The source
correlation $H(X|Y)$ is kept constant and, for each sample
realization of the source, the total rate is progressively increased
beyond the S-W bound, in steps of 0.01 bps, until error-free
decoding is obtained. This operation is repeated on 1000 different
realizations of the source; the mean value and standard deviation of
the rates yielding correct decoding are then computed.

The results have been worked out for block length $N=200$, with
probabilities $p_0=0.5$ and $p_0=0.9$. For  $p_0=0.5$, the
conditional entropy $H(X|Y)$ (i.e., the S-W bound) has been set to
0.5 bps. For  $p_0=0.9$, the joint entropy $H(X,Y)$ has been set to
1 bps; this amounts to coding $Y$ at the ideal rate of $H(Y) \simeq
0.715$ bps, with a S-W bound $H(X|Y) \simeq 0.285$ bps.

The results are reported in Tab. \ref{tab:vrate}. As can be seen, the DAC has a
rate loss of about 0.06 bps with respect to the S-W bound for both the
symmetric and skewed source. The turbo code exhibits a loss of about 0.2 bps
and 0.13 bps. The LDPC-R code has a relatively small loss, i.e., 0.12 bps in
the symmetric case and 0.10 in the skewed one. The LDPC-I code has a slightly
smaller loss, i.e., 0.09 bps in the symmetric case and 0.075 in the skewed one.
However, the DAC still performs slightly better. It should be noted that, while
for LDPC and turbo codes the encoding is done only once thanks to
rate-compatibility, for the DAC multiple encodings are necessary, leading to
higher complexity.

\begin{table}[htb]
\caption{Performance comparison for variable-rate coding: mean and standard
deviation of rate needed for lossless compression.}
\begin{center}
\begin{tabular}{|c||c|c|c|c|}  \hline
& \multicolumn{2}{|c|}{$p_0=0.5$} & \multicolumn{2}{|c|}{$p_0=0.9$}\\
\hline $H(X|Y)$, $H(X,Y)$ & \multicolumn{2}{|c|}{0.50, 1.50} &
\multicolumn{2}{|c|}{0.285, 1.0}
\\ \hline
& mean & st.dev. & mean & st.dev.\\
DAC $N=200$ & 0.56 & 0.04 & 0.32 & 0.03\\
LDPC-R $N=396$ & 0.62 & 0.06 & 0.37 &0.05\\
LDPC-I $N=396$ & 0.59 & 0.06 & 0.35 & 0.05\\
TC16S $N=200$ & 0.71 & 0.11 & 0.42 & 0.08\\
TC16S $N=1000$ & 0.70 & 0.05 & 0.41 & 0.04\\
\hline

\end{tabular}
\end{center}
\label{tab:vrate}
\end{table}

\subsection{Performance versus complexity}
\label{sec:perf_compl}

As has been said, the DAC performance is a function of the block
size and especially of the decoder parameter $M$. Tab.
\ref{tab:complexity} reports comparative decoding results of DAC,
turbo and LDPC codes for various values of $M$ and $N$. The
simulations have been made under the same conditions of Fig.
\ref{fig:1}, i.e. $p_0=0.5$, total average rate equal to 1.5 bps,
and fixed rate of $\underline{x}$ equal to 0.5 bps, considering the
case of $H(X|Y)= 0.25$. Tab. \ref{tab:complexity} reports the
residual BER, and the running time in milliseconds, obtained running
the different decoders on a workstation with Pentium IV 3 GHz
processor running Windows XP.

As can be seen, the DAC complexity grows exponentially with $M$.
Increasing $M$ typically improves performance, and the improvement
is larger as $N$ increases. Comparing DAC and turbo codes at
approximately equal computation time, it can be seen that, for
$N=50$ and $N=200$, the DAC performance is significantly better,
while the turbo code outperforms DAC for $N=1000$. For LDPC codes,
the results for $N=396$ can be compared with the DAC for $N=200$. It
can be seen that, with similar computation time, DAC and LDPC codes
have similar performance. The BER yielded by the LDPC code is four
times smaller than that of DAC, although it would increase going
from $N=396$ to $N=200$.

\begin{table}[htb]
\caption{Decoder complexity and performance for DAC, turbo codes and
LDPC codes.}
\begin{center}
\begin{tabular}{|l|l||c|c|}  \hline
Algorithm &  Parameter & BER & Time (ms) \\
 \hline
DAC $N=50$ & $M=64$ & $1.20 \cdot 10^{-2}$ & 2.26\\
DAC $N=50$ & $M=256$ & $4.89 \cdot 10^{-3}$ & 9.64\\
DAC $N=50$ & $M=512$ & $3.49 \cdot 10^{-3}$ & 22.78\\
DAC $N=50$ & $M=1024$ & $2.93 \cdot 10^{-3}$ & 70.72\\
DAC $N=50$ & $M=2048$ & $2.61 \cdot 10^{-3}$ & 284.16\\
TC16S $N=50$ & 15 iterations & $8.60 \cdot 10^{-3}$ & 9.30\\ \hline
DAC $N=200$ & $M=64$ & $3.15 \cdot 10^{-3}$ & 9.77\\
DAC $N=200$ & $M=256$ & $8.53 \cdot 10^{-4}$ & 44.96\\
DAC $N=200$ & $M=512$ & $4.55 \cdot 10^{-4}$ & 119.94\\
DAC $N=200$ & $M=1024$ & $2.80 \cdot 10^{-4}$ & 394.33\\
DAC $N=200$ & $M=2048$ & $2.00 \cdot 10^{-4}$ & 1538.43\\
TC16S $N=200$ & 15 iterations & $2.74 \cdot 10^{-3}$ & 36.37\\
\hline
DAC $N=1000$ & $M=64$ & $5.36 \cdot 10^{-3}$ & 49.78\\
DAC $N=1000$ & $M=256$ & $1.06 \cdot 10^{-3}$ & 251.32\\
DAC $N=1000$ & $M=512$ & $5.25 \cdot 10^{-4}$ & 766.80\\
DAC $N=1000$ & $M=1024$ & $2.84 \cdot 10^{-4}$ & 2864.06\\
DAC $N=1000$ & $M=2048$ & $1.71 \cdot 10^{-4}$ & 11545.94\\
TC16S $N=1000$ & 15 iterations & $1.2 \cdot 10^{-5}$ &
188.11\\
\hline LDPC-R $N=396$ & 100 iterations & $2.27 \cdot
10^{-4}$ & 16.95\\
LDPC-I $N=396$ & 100 iterations & $2.14 \cdot
10^{-4}$ & 20.18\\
 \hline
\end{tabular}
\end{center}
\label{tab:complexity}
\end{table}

\section{Results: symmetric coding}
\label{sec:results_symm}

In the following we provide results for the symmetric DAC. We
consider two sources with balanced ($p_0=0.5$) and unbalanced
($p_0=0.9$) distribution  with arbitrary rate splitting, and use
$M=2048$.

\subsection{Performance evaluation at fixed rate}

For fixed rate, we set the total rate of $\underline{x}$ and
$\underline{y}$ equal to 1.5 bps. We consider two cases of rate
splitting. In the first case the rate is equally split; we choose
$k^X=k^Y$ so as to achieve a rate of 0.75 bps for each source. In
the second case we encode $\underline{x}$ at 0.6 bps and
$\underline{y}$ at 0.9 bps.


The performance of the symmetric DAC is worked out for $N=200$ and
$N=1000$. Since symmetric DSC coders typically reconstructs each
sequence either without any errors or with a large number of errors
\cite{xiong_tit}, we report the frame error rate (FER) instead of
the residual BER, i.e. the probability that a data block contains at
least one error after joint decoding. For each point, we simulated
at least $10^7$ bits.

\begin{figure}[htb]
\centering
\includegraphics[width=12cm,clip]{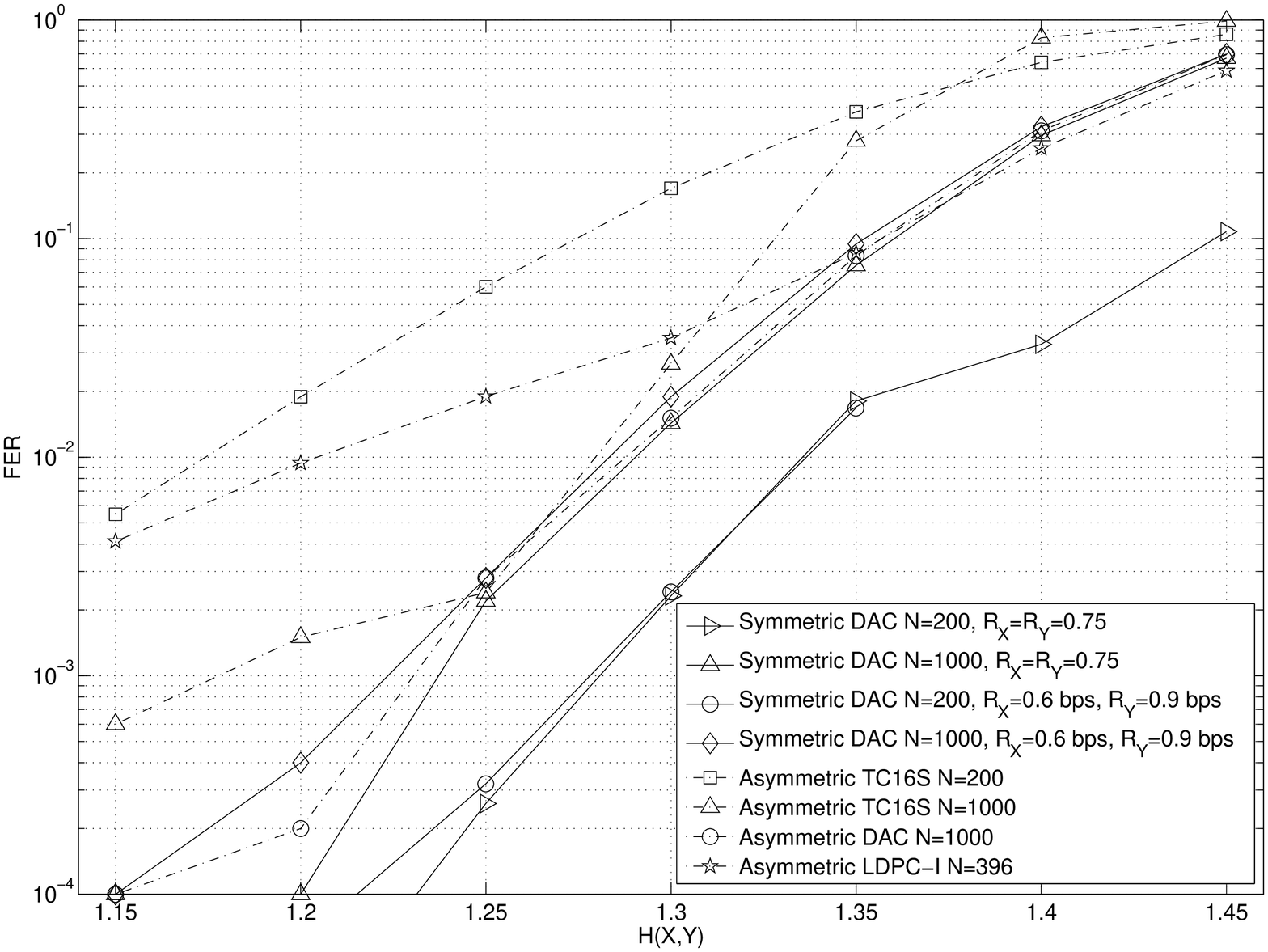}
\caption{Performance comparison of data communication systems
($p_0=0.5$, total rate = 1.5 bps). DAC: distributed arithmetic
coding; TC16S: 16-state turbo code with S-random interleaver;
LDPC-I: irregular LDPC codes from \cite{varodayan}.}
\label{fig:symm_fr}
\end{figure}

Fig. \ref{fig:symm_fr} shows the results for the symmetric DAC.
Comparisons with other algorithms can be done based on the following
remarks. In \cite{frias_sp}, a symmetric S-W coder is proposed
employing turbo codes, which can obtain any rate splitting. In the
case that one source is encoded without ambiguity, this reduces to
the asymmetric turbo-based S-W coder we have employed in Sect.
\ref{sec:results_asymm}. In \cite{frias_sp} it is reported that this
algorithm achieves its best performance in the asymmetric points of
the S-W region, while it is slightly poorer in the intermediate
points. Therefore, in Fig. \ref{fig:symm_fr} we report the FER
corresponding to the best turbo code shown in Fig. \ref{fig:1} for
$N=200$ and $N=1000$, as this lower-bounds the FER achieved by
\cite{frias_sp} over the entire S-W region. Moreover, we also report
the FER achieved by irregular LDPC codes with block length $N=396$
\cite{varodayan}. The asymmetric algorithm in \cite{varodayan} has
been extended in \cite{guillemot_ldpc} to arbitrary rate splitting,
showing that the performance is uniformly good over the entire S-W
region. Finally, we also report the FER curve of the asymmetric DAC
for $N=1000$.

In Fig. \ref{fig:symm_fr}, the results for symmetric coding are very
similar to what has been observed in the asymmetric case. The DAC
achieves very similar BER for $N=200$ and $N=1000$; hence, the FER
is smaller for $N=200$. The results are almost independent of the
rate splitting between $\underline{x}$ and $\underline{y}$, as can
be seen by comparing the two rate-splitting cases as well as the
asymmetric DAC. The turbo codes for $N=200$ and $N=1000$, and the
irregular LDPC code, exhibit poorer performance than DAC.

\subsection{Performance evaluation at variable rate}

\begin{figure}[htb]
\centering
\includegraphics[width=12cm,clip]{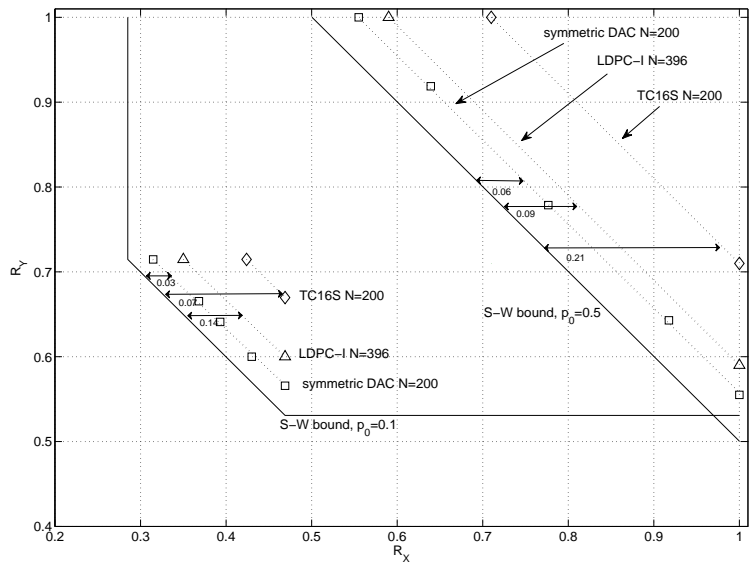}
\caption{Performance comparison at variable rate. The curves in the
top-right corner refer to the case of $p_0=0.5$, and those in the
bottom-left corner to $p_0=0.9$. DAC: distributed arithmetic coding;
TC16S: 16-state turbo code with S-random interleaver; LDPC-I:
irregular LDPC codes from \cite{varodayan}. The solid curves
represent the S-W bound.} \label{fig:symm_vr}
\end{figure}

For variable rate coding, we consider the same two settings as in
Sect. \ref{sec:res_vr}, i.e., block length $N=200$, with
probabilities $p_0=0.5$ and $p_0=0.9$; in the first case the
conditional entropy has been set to 0.5 bps, while in the second
case the joint entropy $H(X,Y)$ has been set to 1 bps. The results
are shown in Fig. \ref{fig:symm_vr}. As can be seen, the performance
of the symmetric DAC is uniformly good over the entire S-W region,
and is significantly better than turbo codes and LDPC codes. In
particular, the DAC suboptimality is between 0.03-0.06 bps, as
opposed to 0.07-0.09 for the irregular LDPC code, and 0.14-0.21 for
the turbo code. It should be noted, however, that variable rate
coding requires feedback, while the S-W bound is achievable with no
feedback, with vanishing error probability as $N\rightarrow \infty$.
In our simulations we re-encode the sequence at increasing rates (in
steps of 0.01 bps), which represents the optimal DAC performance if
the encoder could exactly predict the decoder behavior.

\section{Discussion and conclusions}
\label{sec:concl}

We have proposed DAC as an alternative to existing DSC coders based
on channel codes. DAC can operate in the entire S-W region,
providing both asymmetric and symmetric coding.

DAC achieves good compression performance, with uniformly good
results over the S-W rate region; in particular, its performance is
comparable with or better than that of turbo and LDPC codes at small
and medium block lengths. This is very important in many
applications, e.g., in the multimedia field, where the encoder
partitions the compressed file into small units (e.g., packets in
JPEG 2000, slices and NALUs in H.264/AVC) that have to be coded
independently.

As for encoding complexity, which is of great interest for DSC, DAC
has linear encoding complexity, like a classical AC \cite{helfgott}.
Turbo codes and the LDPC codes in \cite{varodayan} also have linear
encoding complexity, whereas general LDPC codes typically have more
than linear, and typically quadratic complexity
\cite{richardson_urb}. As a consequence, the complexity of DAC is
suitable for DSC applications.

A major advantage of DAC lies in the fact that it can exploit
statistical prior knowledge about the source very easily. This is a
strong asset of AC, which is retained by DAC. Probabilities can be
estimated on-the-fly based on past symbols; context-based models
employing conditional probabilities can also be used, as well as
other models providing the required probabilities. These models
allow to account for the nonstationarity of typical real-world
signals, which is a significant advantage over DSC coders based on
channel codes. In fact, for channel codes, accounting for
time-varying correlations requires to adjust the code rate, which
can only be done for the next data block, incurring a significant
adaptation delay. Moreover, with channel codes it is not easy to
take advantage of prior information; for turbo codes it has been
shown to be possible \cite{frias_jsac}, employing a more
sophisticated decoder.

Another advantage of the proposed DAC lies in the fact that the
encoding process can be seen as a simple extension of the AC
process. As a consequence, it is straightforward to extend an
existing scheme employing AC as final entropy coding stage in order
to provide DSC functionalities.


\end{document}